# Emergence of Correlations in Alternating Twist Quadrilayer Graphene

G. William Burg[1], Eslam Khalaf [2], Yimeng Wang[1], Kenji Watanabe[3], Takashi Taniguchi[4],

Emanuel Tutuc[1]

[1]Microelectronics Research Center, Department of Electrical and Computer Engineering, The University of Texas at Austin, Austin, Texas 78758, USA
[2]Department of Physics, Harvard University, Cambridge, MA 02138
[3]Research Center for Functional Materials, National Institute of Materials Science, 1-1 Namiki Tsukuba, Ibaraki 305-0044, Japan
[4]International Center for Materials Nanoarchitectonics, National Institute of Materials Science, 1-1 Namiki Tsukuba, Ibaraki 305-0044, Japan

Corresponding author: Emanuel Tutuc, etutuc@mail.utexas.edu

**Recently, alternating twist multilayer graphene (ATMG) has emerged as a family of moiré systems that share several fundamental properties with twisted bilayer graphene, and are expected to host similarly strong electron-electron interactions near the magic angle. Here, we study alternating twist quadrilayer graphene (ATQG) samples with twist angles of 1.96° and 1.52°, which are slightly removed from the magic angle of 1.68°. At the larger angle, we find signatures of correlated insulators only when the ATQG is hole doped, and no signatures of superconductivity, and for the smaller angle we find evidence of superconductivity, while signs of the correlated insulators weaken. Our results provide insight into the twist angle dependence of correlated phases in ATMG and shed light on the nature of correlations in the intermediate coupling regime at the edge of the magic angle range where dispersion and interaction are of the same order.**

The emergence of two-dimensional van der Waals materials has enabled a new field of study in which moiré superlattices are engineered by stacking multiple van der Waals materials and applying a controlled twist between layers. For certain moiré heterostructures, there are a series of 'magic angles' at which the lowest lying energy bands of the moiré band structure become extremely flat, enabling the Coulomb interaction strength to greatly exceed the kinetic energy of electrons in the band, favoring electron-electron correlations[1]. Starting with the discovery of correlated insulators and superconductivity within the flat bands of magic angle twisted bilayer graphene (MATBG)[2,3], substantial effort has been made to better understand the fundamental physics of moiré flat bands, and to expand the family of moiré heterostructures. Recent studies have revealed additional correlation-driven states in MATBG, such as ferromagnetism[4,5] and Chern insulators[6–9], while others have focused on new types of twisted bilayers consisting of monolayer-bilayer graphene[10–12], bilayer-bilayer graphene[13–17], trilayer graphene on hBN[18,19], and transition metal dichalcogenides[20–22]. These systems exhibit similar correlated phases as in TBG but they differ in some fundamental aspects related to symmetry and band topology. In addition, they show no or rather weak signatures of superconductivity in contrast to TBG where robust superconductivity has been observed.

One family of moiré systems which share most of the TBG properties is alternating twist multilayer graphene (ATMG), a family of graphene moiré patterns consisting of $m \geq 3$ graphene monolayers with a twist angle ($\theta$) that alternates between $+\theta$ and $-\theta$ between each successive pair of layers[23]. The ATMG band structure can be exactly decomposed into $m/2$ distinct TBG band structures, each at an effective twist angle different than $\theta$, for even $m$, or $(m-1)/2$ TBG band structures plus a Dirac cone for odd $m$. At the $m$-dependent magic angle, ATMG hosts flat bands similar to those of MATBG concurrently with dispersive bands, and for $m \geq 5$ can host multiple coexisting flat bands[23,24]. Realizing these systems requires overcoming the technical challenge of controlling several twist angles at once.

The simplest member of the ATMG family is ATTG, consisting of three graphene layers with relative twist angles $+\theta$ and $-\theta$. ATTG maps to a TBG with the interlayer coupling scaled by a factor of $\sqrt{2}$ in addition to a dispersive Dirac band. Several recent studies have examined ATTG at the magic angle, revealing robust electric field-tunable superconductivity that persists across a wide range of densities[25–29]. However, the interplay between superconductivity and correlated insulators in ATMG, and their dependence on $\theta$ remains unexplored. Here, we report the fabrication and characterization of ATQG close to, but slightly removed from the magic angle of 1.68°. We focus on two samples with $\theta = 1.96°$ and 1.52°. We find that above the magic angle, resistance maxima suggestive of correlated insulators appear at half filling of the hole-doped side of the moiré flat bands, while superconductivity is absent. Note that, away from neutrality there is always residual conductance from the coexisting dispersive bands. Below the magic angle we observe the opposite trend, wherein evidence of superconductivity is present, while the correlated insulators weaken. In both samples a small perpendicular magnetic field enhances the correlated insulating features, and quickly suppresses nascent superconductivity when present. In contrast, an in-plane magnetic field has little effect on the correlated insulators, and evidence of superconductivity persists in fields up to 7 T. A transverse electric field appears to enhance both the correlated insulators in the large angle sample and the superconductivity in the small angle sample.

Figure 1a shows a schematic of ATQG, consisting of four graphene monolayers with an alternating twist angle between each layer pair. Figures 1b-d show the band structure of ATQG, taking into account relaxation[23,24], at the two sample angles and the magic angle, 1.68°. At the magic angle, the band structure can be separated into two TBG subsystems: one TBG at the first magic angle, giving rise to a pair of flat bands (blue lines in Fig. 1c), and another TBG away from magic angle, which contributes two dispersive bands (green lines in Fig. 1c). At the two sample angles, the flat bands become more dispersive, but the overall character of the band structure is similar. Figure 1e shows the longitudinal resistance ($R_{xx}$) contour plot as a function of carrier density ($n$) and transverse electrical field ($E$) measured in the $\theta = 1.96°$ sample at a temperature $T = 160$ mK.

We observe $R_{xx}$ maxima at charge neutrality and at $n = -8.2 \times 10^{12}$ cm$^{-2}$ and $-4.9 \times 10^{12}$ cm$^{-2}$, which correspond to full filling and half filling of the flat valence band, respectively. While in the single particle picture, a gap at neutrality is not expected due to the semi-metallic dispersive bands, an interaction-generated gap in the flat band can be induced in the dispersive bands, which, unlike in ATTG, occurs because the two TBG subsystems are not distinguished by symmetry[24]. The full filling density is denoted as $n_s$ and is related to $\theta$ by $n_s \cong 8\theta^2/\sqrt{3}a^2$, where $\theta$ is in radians and $a = 2.46$ Å is the graphene lattice constant. The Fig. 1e upper $x$-axis shows $n$ normalized to $n_s$. For $E$ near zero, the resistance at $n = -n_s/2$ and $-n_s$ is small due to residual conductance from the dispersive bands, but reaches a maximum as a gap opening is induced in the dispersive bands with increasing $E$[24]. At large $E$, no resistance maxima are observed at half filling, consistent with calculations that show the two TBG subsystems hybridize and increase the bandwidth of the flat band. The appearance of correlated insulators exclusively in the valence band is in contrast with what is typically observed in TBG, in which correlated insulators are usually more robust when filling the conduction band, although there is some evidence that correlated insulators become stronger on the hole side at larger than magic angles[30,31].

Figure 1f shows the temperature dependence of the maximum $R_{xx}$ at $n = 0$, $-n_s/2$ and $-n_s$. For each dataset, $R_{xx}$ shows an upward trend at low $T$, suggestive of insulating behavior (see also Extended Data Fig. 1). Figure 1g shows $R_{xx}$ vs $n/n_s$ and $T$ along the dotted line in Fig. 1e. At high $T$, only the band insulators are visible, while below 15 K the correlated insulator begins to appear and regions of low resistance develop on either side of it.

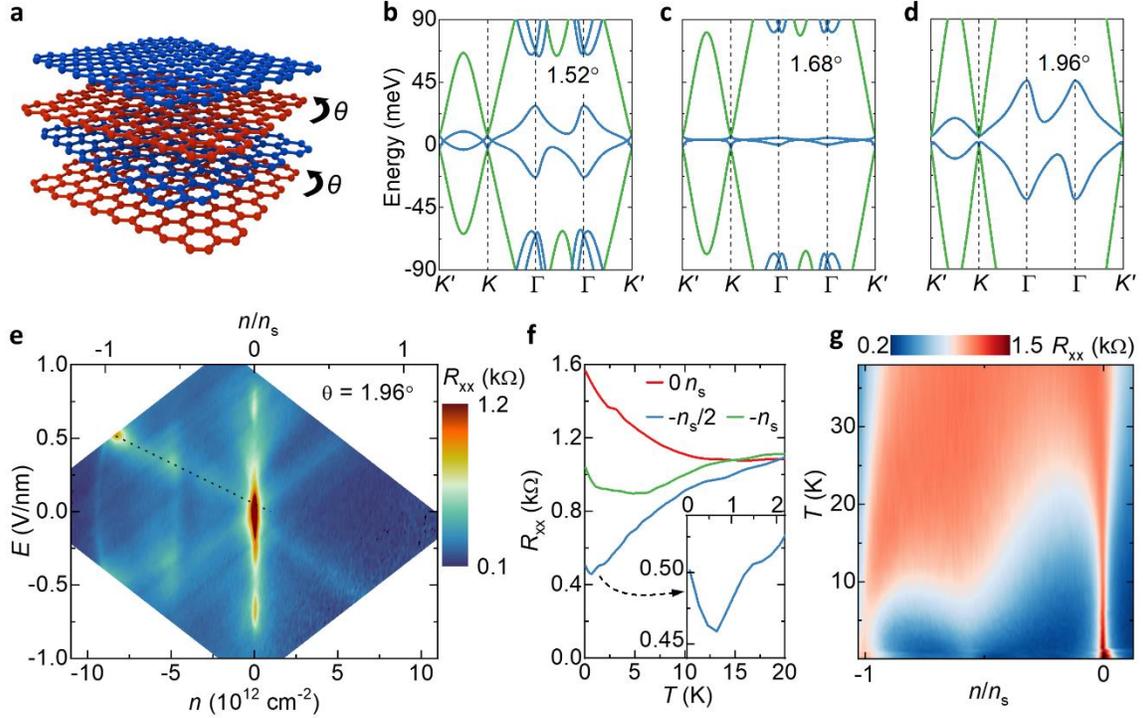

**Fig. 1. Transport and temperature dependence in ATQG. a**, Schematic of ATQG, with the second and fourth graphene monolayers (red) rotated by $\theta$ relative to the first and third monolayers (blue). **b-d**, Band structure of ATQG at $\theta = 1.52°$, $1.68°$, and $1.96°$, respectively, displaying the flat bands of the TBG subsystem near magic angle (blue), and the dispersive bands of the TBG subsystem away from magic angle (green). **e**, $R_{xx}$ vs. $n$ and $E$ measured in the $\theta = 1.96°$ sample at $T = 160$ mK. The top x-axis shows $n$ normalized to $n_s$. Correlated insulators appear when the valence band is half filled, in small ranges of $E$ values. **f**, $R_{xx}$ vs $T$ measured at $n = 0$, $-n_s/2$, and $-n_s$ along the dotted line in panel **e**. The inset shows a zoomed in view of the $-n_s/2$ resistance below 2 K. **g**, $R_{xx}$ vs. $n/n_s$ and $T$ along the dotted line in **e**.

To further elucidate the nature of the correlated insulators, we examine the effect of an applied magnetic field. Figures 2a and b show $R_{xx}$ and the Hall resistance ($R_{xy}$), respectively, as a function of $n/n_s$ and $E$ in a perpendicular field $B = 1$ T. At this field, the $R_{xx}$ maxima of the correlated and band insulators extend to a wider range of $E$-fields relative to $B = 0$ T and appear on the electron side as well. At $n = -n_s/2$ and between $E = \pm 0.5$ V/nm, the $R_{xy}$ data show an abrupt change in value away from zero, a feature also observed in MATBG[2,32] and ATTG[25] and associated with a phase transition at the correlated insulator[31,33]. At larger $E$, $R_{xy}$ transitions through zero smoothly at $n = -n_s/2$, as expected when filling a single-particle band, suggesting a suppression of correlations and consistent with the disappearance of $R_{xx}$ maxima. A similar behavior is observed at $n = +n_s/2$, although in general signatures of the correlated insulator are weaker on the electron side.

An interesting feature of the $B = 1$ T data is the expansion of the half-filling resistance maxima to a larger range of $E$-fields compared to the zero $B$-field data (see also Extended Data Fig. 2). Figure 2c shows $R_{xx}$ at $n = -n_s/2$ as a function of $E$ and $B$. For small $B$, the $R_{xx}$ maxima are confined to small regions near $|E| = 0.3$ V/nm, but quickly expand and merge by $B = 0.5$ T. The applied $B$-field induces gaps in the dispersive band through the formation of Landau levels. For small $B$, a combined $E$ and $B$-field are required to sufficiently open the band gap to reveal the correlated insulators but above a critical $B$, the correlated insulators are visible for all small $E$. These data support the scenario where correlated insulators are most likely to emerge if the flat bands are spectrally isolated from neighboring bands[13]. Figure 2d shows $R_{xx}$ vs. $n/n_s$ at $E = 0$, for

$B$-fields ranging from 0 T to 0.5 T, demonstrating the marked increase in the correlated and full-filling insulator resistances for small $B$-fields.

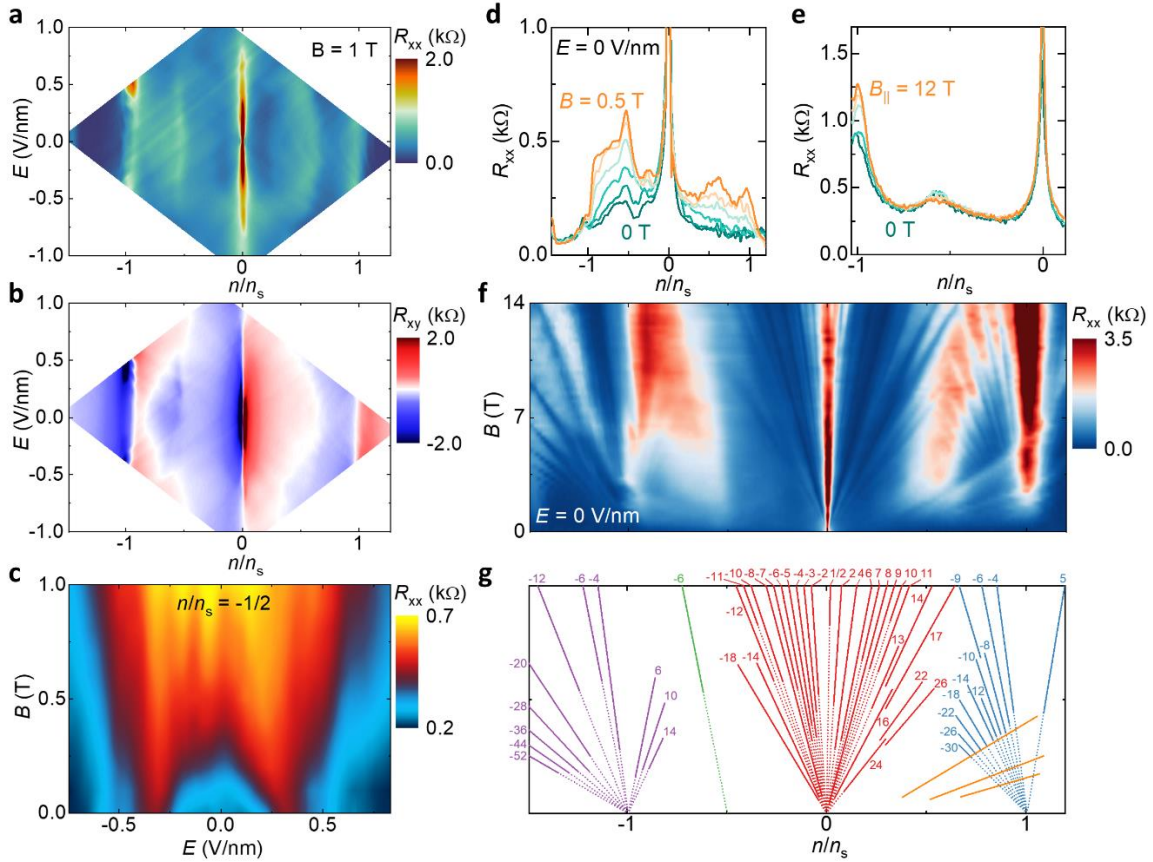

**Fig. 2.** $\theta = 1.96°$ **ATQG in a magnetic field. a, b** $R_{xx}$ (**a**) and $R_{xy}$ (**b**) vs. $n/n_s$ and $E$ at $B = 1$ T and $T = 160$ mK. **c**, $R_{xx}$ vs. $E$ and $B$ along a vertical slice of **a** at $n/n_s = -1/2$. Resistance maxima that are initially localized to small regions of $E$ expand and merge with increasing $B$. **d**, $R_{xx}$ vs. $n/n_s$ at $E = 0$ V/nm as a function of $B$ between 0 and 0.5 T, in increments of 0.1 T, showing correlated and full-filling insulators respond strongly to a small $B$-field. **e**, $R_{xx}$ vs. $n/n_s$ along the dotted line in Fig. 1c as a function of $B_{\parallel}$ between 0 and 12 T, in increments of 4 T. The correlated insulator shows little dependence on $B_{\parallel}$. **f**, $R_{xx}$ vs. $n/n_s$ and $B$ at $E = 0$ V/nm, which shows oscillations associated with the formation of Landau levels originating from correlated and band insulators. **g**, Fits and labels of individual Landau levels in **f**.

We also consider the effect of an in-plane magnetic field ($B_{\parallel}$). In the limit of zero layer thickness, an in-plane magnetic field only couples to the spin degree of freedom via the Zeeman effect, but because of the finite layer thickness of the electron system considered here, $B_{\parallel}$ can also couple via an orbital effect[24,34]. The orbital coupling can be understood through a layer-dependent vector potential $d$ ($l\,\hat{z} \times \boldsymbol{B}_{\parallel}$), where $l = 1, 2, 3, 4$ is the layer index and $d = 3.4$ Å is the interlayer distance. The resulting orbital coupling is obtained by modifying the kinetic component of the Hamiltonian in layer $l$ as $h_l = v_F \sigma \cdot \nabla \rightarrow v_F \sigma \cdot \nabla - g_{orb} \mu_B \sigma \cdot (l\,\hat{z} \times \boldsymbol{B}_{\parallel})$ where $g_{orb} = ev_F d/\mu_B \sim 6$ is the orbital g-factor, which is of the same order as Zeeman splitting; $e$ is the electron charge, $v_F$ is the Fermi velocity of monolayer graphene, and $\mu_B$ is the Bohr magneton. For ATQG, a recent theoretical study has shown that the effective orbital coupling in the TBG subsystem hosting the flat band has an orbital coupling that is smaller by a factor of 20 compared to TBG[24], leading to a reduced sensitivity of correlated states to $B_{\parallel}$. Figure 2e shows $R_{xx}$ vs. $n/n_s$ at $B_{\parallel}$ from 0 T to 12 T. The correlated insulator at $n = -n_s/2$ undergoes a slight broadening and reduction of the resistance peak up to $B_\parallel = 12$ T, which is consistent with the prior discussion and indicative of a spin unpolarized state, as a Zeeman gap should increase with $B_\parallel$.

Figure 2f shows $R_{xx}$ as a function of $n/n_s$ and $B$ at a constant $E = 0$ V/nm, and exhibits quantum oscillations arising from the formation of Landau levels in a perpendicular magnetic field (see Methods). We observe Landau fans emerging from $n/n_s = -1, -1/2, 0, +1$, with the filling factor ($\nu$) of individual levels labeled and the sub-bands indicated by different colors in Fig. 2g. The fan at $n/n_s = 0$ shows single integer Landau levels, signaling a full lifting of the four-fold spin and valley degeneracy, and is indicative of high sample quality. Furthermore, the robust fans at $n = \pm n_s$ suggest good angle uniformity. At $n = -n_s/2$, there is a single Landau level corresponding to $\nu = -6$, which is expected to occur for a band with Chern number -2, and a contribution of -4 from the zeroth Landau level of the dispersive band. The emergence of a Chern –2 band at $n = -n_s/2$ at finite $B$-field is also present in MATBG[8,35]. We observe an additional fan at positive $n$ and smaller $B$-fields with much lower slopes than the other fans, which we attribute to the partial filling of the dispersive band, as also seen in ATTG[25,26].

We now consider the ATQG sample with $\theta = 1.52°$, that is closer to but smaller than the 1.68° magic angle. Figure 3a shows $R_{xx}$ vs. $n/n_s$ and $E$ at $T = 160$ mK. The data show broad, $E$-dependent resistance maxima near $\pm n_s/2$ and $\pm n_s$, as well as several regions of low resistance at $\pm(n_s/2 + \delta)$, $\pm(n_s + \delta)$, and $0 - \delta$, where $\delta$ represents a small density value. It is instructive to examine these regions more closely, in particular near half filling, where superconductivity most commonly occurs in TBG. Figures 3b-e show the four-point differential resistance (d$V$/d$I$) vs. current ($I$) at four representative points labeled by the colored arrows in Fig. 3a. In the regions near charge neutrality (Fig. 3c) and full filling (Fig. 3e) the differential resistance remains insensitive to the current. Conversely, near half filling (Fig. 3b, d) the d$V$/d$I$ data show a pronounced critical current behavior, characterized by sharp peaks in d$V$/d$I$ at the transition from a low to high resistance state. The regions that exhibit the strongest critical current behavior occur at finite $E$-fields (see Extended Data Fig. 3), consistent with similar observations in ATTG, and in agreement with calculations suggesting an increase in the critical temperature with $E$ for twist angles close to but smaller than the magic angle[24].

Figure 3f shows the temperature dependence of $R_{xx}$ vs. $n/n_s$ at a constant $E = -0.44$ V/nm. At densities near $n_s/2 + \delta$, a clear dome of low resistance forms while little temperature dependence is observed for all other $n/n_s$, including the low resistance regions near charge neutrality and full filling. Figure 3g shows the temperature dependence of $R_{xx}$ at the optimal doping of Fig. 3f (orange arrow in Fig. 3a) over a larger temperature range. Around 2 K, the resistance drops sharply, but plateaus at a finite resistance of ~350 Ω for temperatures below 1 K. Using 50% of the normal state resistance as the criteria for the critical temperature ($T_c$) gives $T_c \approx 1.34$ K. The inset of Fig. 3g shows d$V$/d$I$ vs. $I$ on logarithmic scales. The temperature at which the d$V$/d$I$ characteristic near the critical current is proportional to $I^2$ (dashed line) gives the Berezinskii-Kosterlitz-Thouless transition temperature ($T_{BTK}$) and, to best fit, the data yield $T_{BTK} \approx 1.29$ K.

Overall, the d$V$/d$I$ vs. $I$ and temperature dependence data are strongly suggestive of nascent superconductivity (see also Extended Data Figs. 4 and 5). However, given the finite resistance at the lowest temperatures, it is best to exercise diligence. Joule heating has been proposed as a mechanism to produce non-linear current-voltage characteristics and strong temperature dependence, particularly in twisted double bilayer graphene[17]. In these cases, the data consistently show gradual transitions at a finite current leading to broad peaks or no peaks in the d$V$/d$I$ vs. $I$[16,17,19,21,36], in contrast to the sharp d$V$/d$I$ peaks observed here. A second normal-state mechanism that can produce abrupt transitions in the d$V$/d$I$ is non-equilibrium transport[37], where velocity saturation leads to an apparent critical current behavior. However, these characteristics arise at very large current densities on the order of 50 µA/µm, several orders of magnitude larger than the currents used here (~100 nA/µm). Furthermore, a critical current is observed at all densities away from neutrality in the non-equilibrium regime, and not exclusively near fractional filling of the bands, as we observe (Figs. 3b-e).

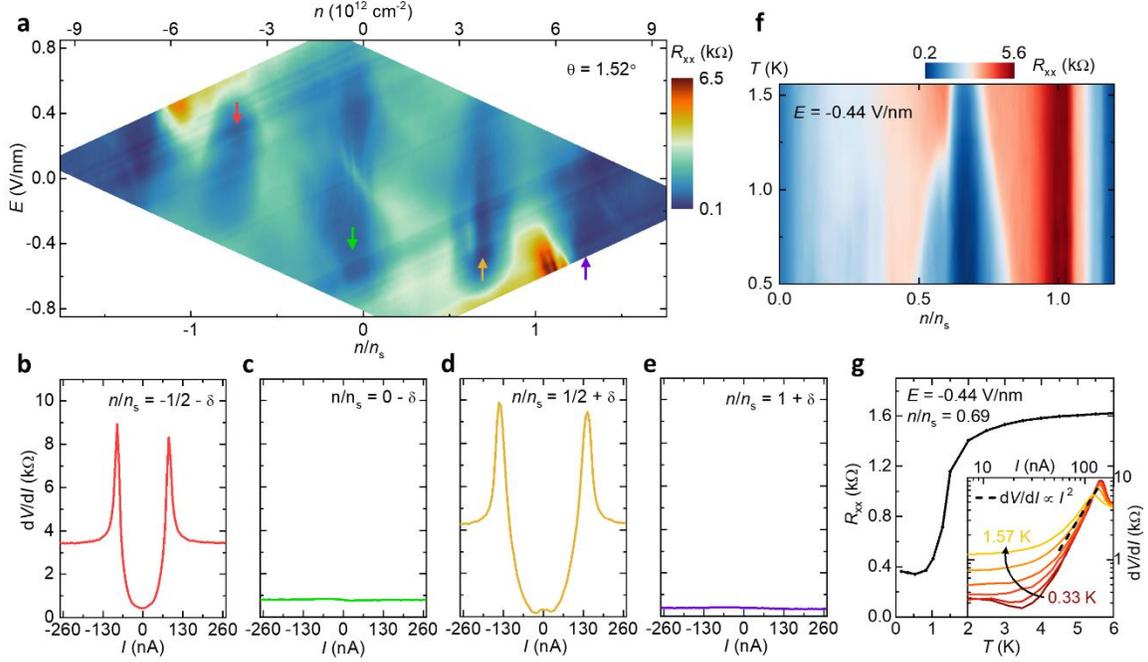

**Fig. 3. Evidence of superconductivity in $\theta = 1.52°$ ATQG. a**, $R_{xx}$ vs. $n/n_s$ and $E$ at $T = 160$ mK. The top $x$-axis shows the total density, $n$. Multiple areas of low resistance appear near charge neutrality, and half filling and full filling of the flat bands. **b-e**, $dV/dI$ vs. $I$ at four representative $n$ and $E$ as labeled by the corresponding color arrows in **a**. Regions near $\pm n_s/2$ show critical current characteristic while other low resistance regions do not. **f**, $R_{xx}$ vs. $n$ and $T$ along $E = -0.44$ V/nm, showing a dome of low resistance at $n_s/2 + \delta$. **g**, $R_{xx}$ as a function of $T$ at $n/n_s = 0.69$ and $E = -0.44$ V/nm, illustrating the abrupt resistance decrease at 2 K. The inset shows $dV/dI$ vs. $I$ on a logarithmic scale at different temperatures and the black dashed line shows a line of $dV/dI \propto I^2$, which is used to determine $T_{BTK}$.

A key test of a superconducting state is the response to a small perpendicular $B$-field, which has been used in MATBG to identify states with finite resistance at the lowest temperatures[5]. The Joule heating effect has not been considered in a magnetic field[17] and the critical current that arises from non-equilibrium transport was shown to be insensitive to a magnetic field up to 2 T[37].

Figure 4a shows $dV/dI$ as a function of $I$ and $B$ at $T = 160$ mK at the same $E$ and $n$ values as in Figs. 3d and 3g. The critical current is strongly dependent on the applied $B$-field, steadily decreasing with increasing field, and is fully extinct at ~350 mT, consistent with a nascent superconducting state. Figure 4b shows the corresponding $I$ vs. $V$ data at 100 mT intervals, demonstrating the marked reduction of the switching behavior at the critical current as the characteristic trends towards linear with $B$-field. The Fig. 4a and 4b data give an approximate critical $B$-field ($B_c$) of 250 mT, which can be used to estimate the Ginzburg-Landau coherence length ($\xi$) with $B_c = [\Phi_0/(2\pi\xi)](1 - T/T_c)$, where $\Phi_0 = h/(2e)$ is the superconducting flux quantum; $h$ is the Planck constant. Using the given relation, we find $\xi \approx 34$ nm. We do not observe Fraunhofer oscillations in the $B$-field data, which are often seen in MATBG[3,30,32], and occur due to phase coherent transport through small normal regions – that arise from charge inhomogeneity – interspersed within the superconducting bulk. It is possible that in this sample only isolated pockets of the channel are superconducting and the normal regions separating them are too large to maintain phase coherence, which may also explain the non-zero minimum resistance.

In contrast to a perpendicular magnetic field, in-plane magnetic fields are expected to have little effect on superconducting states in all $m$-layer ATMG. Odd-layer ATMG is explicitly protected from pair-breaking in an in-plane field due to its mirror and time reversal symmetry[35], and has been demonstrated experimentally in ATTG in which superconductivity survives significantly beyond the Pauli limit[38]. Even-layer ATMG lacks mirror symmetry and should respond to $B_\parallel$ via orbital coupling in the same way as MATBG. The orbital effect in TBG is obtained by projecting the modified Hamiltonian onto the flat bands which yields a momentum-dependent correction to the dispersion that has opposite sign between the two valleys, leading to pair-breaking for inter-valley superconductivity at a field that depends on both doping and field direction but is roughly the same as the Pauli field[24,39,40]. However, this effect in ATQG is expected to be scaled down by a factor of 20[24]. To better illustrate the reduced effect of an in-plane $B$-field, we first calculate the theoretical Pauli limit ($B_P$), assuming a weak coupling superconductor and a $g$-factor of 2, using $B_P = 1.86$ [T/K] $T_c$, which for our sample yields $B_P \approx 2.5$ T. Figure 4c shows $R_{xx}$ vs. $I$ and $B_\parallel$ at multiple temperatures and, indeed, the evidence of superconductivity persists up to a critical field of roughly 7 T at the lowest temperature, well beyond the expected Pauli limit and almost an order of magnitude larger than in MATBG[40]. With increasing temperature, the critical field and critical current both decrease, and at $T = 2$ K no signs of superconductivity are observed, consistent with the Fig. 3g data. Furthermore, we note the Fig. 4a and 4c data were taken using different pairs of contacts and that in general the observed transport characteristics are consistent across multiple sets of contacts (see Extended Data Fig. 6).

The data presented here demonstrate the existence of correlations in ATQG, even when removed from the exact magic angle. While the physics of ATQG is closely analogous to TBG, there are distinct similarities and differences as a function of $\theta$. A comparison between our ATQG samples and TBG can be made by dividing the ATQG twist angle by the golden ratio, ≈1.62, to find an equivalent TBG twist angle. The $\theta = 1.96°$ and $\theta = 1.52°$ samples map to TBGs with $\theta = 1.21°$ and $\theta = 0.94°$, respectively. The appearance of stronger insulating states on the hole side compared to the electron side in the large angle sample is similar to earlier results in TBG[31], where correlated insulators probed through electronic compressibility measurements become stronger on the hole side for angles approaching 1.20°. In addition, the observation of superconductivity without correlated insulators for angles as low as 0.79° has been reported in TBG[41]. Our observations depart from the TBG picture when considering the asymmetry in the twist angle dependence of the superconductivity and correlated insulators, where the former is more prominent at smaller angles and the latter is more prominent at larger angles, while in TBG, superconductivity is more robust upon moving away from the magic angle in both directions[30,41,42]. While more data is needed to confirm this as a distinct feature of ATQG, it suggests that the presence of the dispersive bands may play a role in the stability of different correlated phases.

Our results also demonstrate the robustness of the superconducting phase to an in-plane magnetic field, exhibiting large Pauli limit violation similar to ATTG. This strongly suggests that superconductivity in both systems is not spin-singlet. Additionally, the insensitivity of the insulating features to an in-plane field indicates a non-spin polarized parent state. Both observations point towards a spin-valley locked state as the most likely parent state which yields a superconducting order that consists of a mixture of spin-singlet and triplet[43,44]. Our findings show that ATQG, and ATMG more generally, is a unique playground for probing interactions beyond TBG. It highlights a decoupling of the superconducting phase and correlated insulators with twist angle, as well as a robustness to an in-plane magnetic field that is absent in TBG.

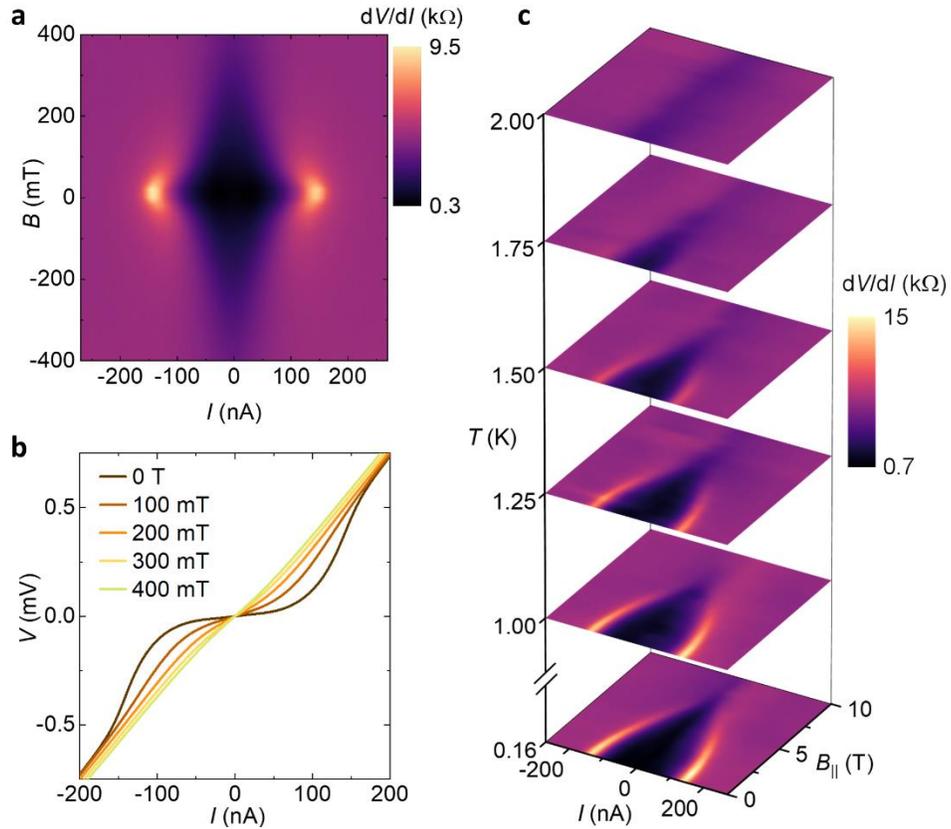

**Fig. 4. Magnetic field response of nascent superconductivity. a**, d$V$/d$I$ vs. $I$ and $B$ at $n/n_s$ = 0.69 and $E$ = -0.44 V/nm. The critical current is quickly suppressed by a small $B$-field. **b**, $V$ vs. $I$ as a function of $B$ between 0 and 400 mT, in 100 mT steps. **c**, d$V$/d$I$ vs. $I$ and $B_\parallel$ at multiple temperatures. At the lowest temperature, the critical field is $B_\parallel \approx 7$ T. The critical current and field both shrink with increasing $T$ and are entirely absent at 2 K.

**Acknowledgements**

The work at The University of Texas at Austin was supported by the National Science Foundation (NSF) grants MRSECDMR-1720595, and EECS-2122476; Army Research Office under grant no. W911NF-17-1-0312; and the Welch Foundation grant F-2018-20190330. Work was partly done at the Texas Nanofabrication Facility supported by the NSF grant no. NNCI-2025227. K.W. and T.T. acknowledge support from the Elemental Strategy Initiative conducted by the MEXT, Japan (grant no. JPMXP0112101001), and JSPS KAKENHI (grant nos. JP19H05790 and JP20H00354).

**Author Contributions**

G. W. B. and E. T. conceived and designed the experiment. G. W. B. fabricated and measured the samples, and Y. W. assisted with measurements. G. W. B., E. K., and E. T. analyzed the data. E. K. provided band structure calculations. K. W. and T. T. provided the boron nitride crystals. All authors contributed to discussions and writing of the manuscript.

**Competing Interests**

The authors declare no competing interests.

*Note added.* – during the preparation of this manuscript we became aware of two related studies, which show robust superconductivity in ATQG at the magic angle[45,46].

## Methods
### Sample Fabrication

To prepare ATQG samples, hBN and graphene are separately exfoliated onto Si/SiO$_2$ substrates. Large area graphene flakes, typically with a length of at least 80 µm, are identified using optical microscopy and the single atomic layer thickness is confirmed with Raman spectroscopy. The hBN flakes are similarly identified and their thickness and surface roughness are determined with atomic force microscopy (AFM). The four constituent graphene layers are predefined by cutting the graphene flake using an AFM tip in contact mode. A hemispherical polymer stamp consisting of polydimethylsiloxane (PDMS) and a thin layer of spin-coated polypropylene carbonate (PPC) on a glass slide is used to first selectively pick up an hBN flake with a thickness of approximately 30 nm. Then, individual graphene layers are picked up sequentially by the top hBN, while alternating the twist angle between each successive layer. The target angles during this process are usually ~0.3° larger than the final desired angle, to allow for relaxation during further processing. All pick-ups are performed at a temperature of 60° C. The stack is then released onto a graphite/hBN bottom gate at 170° C, and a graphite top gate is subsequently transferred onto the heterostructure using a similar procedure as described. Finally, electron beam lithography (EBL) and plasma etching are used to define the Hall bar geometry and metal edge contacts are evaporated onto the sample using EBL and physical vapor deposition. The optical micrographs of the samples presented in the main text are shown in Extended Data Fig. 6.

### Measurement Setup

Four-point transport measurements are performed using a low frequency lock-in amplifier, with an excitation current ranging from 1-10 nA. The charge density ($n$) and transverse electric field ($E$) are independently controlled using top ($V_{TG}$) and bottom ($V_{BG}$) gate voltages. The density is given by $n = (V_{TG}C_{TG} + V_{BG}C_{BG})/e$, where $C_{TG}$ and $C_{BG}$ are the top and bottom gate capacitances, respectively, and $e$ is the electron charge. The electric field is given by $E = (V_{TG}C_{TG} - V_{BG}C_{BG})/2\varepsilon_0$, where $\varepsilon_0$ is the vacuum permittivity.

Measurements are performed in two different refrigerators. A dilution refrigerator is used for measurements performed at temperatures between 160 mK and 1.5 K, and a pumped $^4$He refrigerator is used for measurements at temperatures of 1.5 K and higher.

**Assignment of Full Filing Density**

The value of $n_s$ is experimentally determined by fitting Landau fans at large perpendicular magnetic fields (Fig. 2g and Extended Data Fig. 7). At high magnetic fields, the Landau level gaps in the dispersive bands are sufficiently large that the gate-induced charge density predominantly fills the flat bands and $n_s$ can be accurately determined. At zero magnetic field, there is a partitioning of carriers between the flat and dispersive band, and consequently the half and full filling resistance maxima in Fig. 1e and Fig. 3a do not exactly align with the designated $-n_s/2$ and $-n_s$ values on the respective $x$-axes.

**Quantum Oscillations in a Perpendicular Magnetic Field**

In a perpendicular magnetic field ($B$), electrons take on discrete cyclotron orbits known as Landau levels with quantized energies that are proportional to $B$ and have a level index, or filling factor, $v$. Each sub-band ($s$) of the band structure splits into Landau levels, which manifests in transport as minima in the longitudinal resistance every time the charge density ($n$) is modulated to fill a Landau level. Landau levels are indexed by $s$ and $v$, and are related by $n = v \cdot (eB/h) + s \cdot n_s$, where $e$ is the electron charge and $h$ is Planck's constant. Experimentally, these appears as 'fans' of low resistance that extend outward from band or correlated insulators (Fig. 2 and Extended Data Figs. 7 and 8).

**Data Availability**

All supporting data are available from the corresponding author upon reasonable request.

**Code Availability**

The supporting code for this paper is available from the corresponding author upon reasonable request.

**Extended Data Figures:**

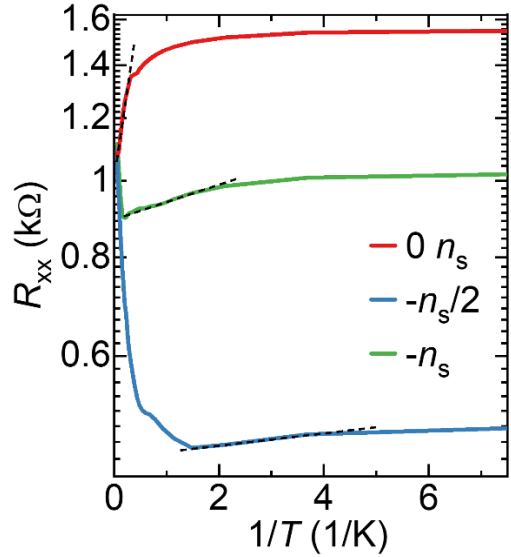

**Extended Data Fig. 1. Arrhenius plot of correlated and band insulators in $\theta = 1.96°$ ATQG.** Log-scale $R_{xx}$ vs. $1/T$ at $n/n_s = 0$, $1/2$, and $1$, using the same data as in Fig. 1f of the main text. Linear portions of each trace are marked by a dashed line.

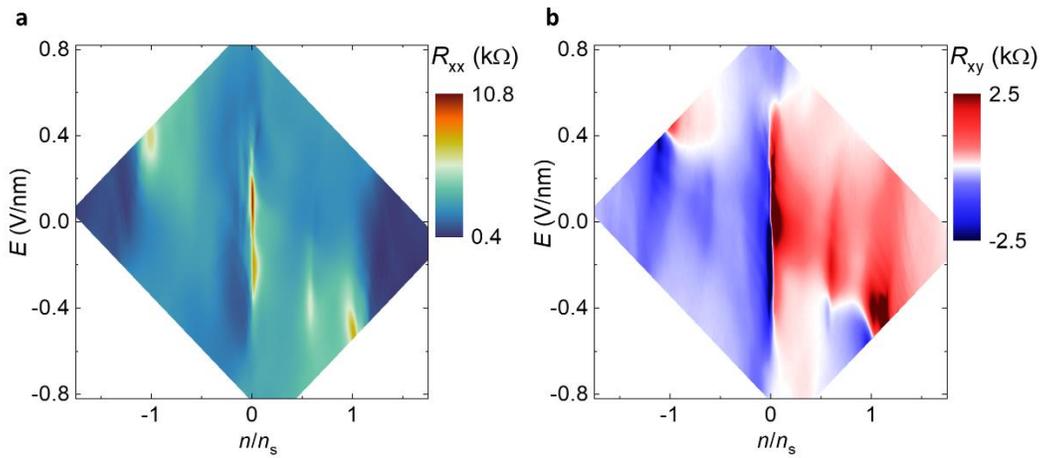

**Extended Data Fig. 2. $\theta = 1.52°$ ATQG at $B = 1$ T. a-b**, $R_{xx}$ (**a**) and $R_{xy}$ (**b**) vs. $n/n_s$ and $E$ at $B = 1$ T. Similar to the Fig. 2 data for $\theta = 1.96°$ ATQG, $R_{xx}$ maxima become more apparent at half and full filling and $R_{xy}$ data show trend reversals indicative of a phase transition at the correlated insulators.

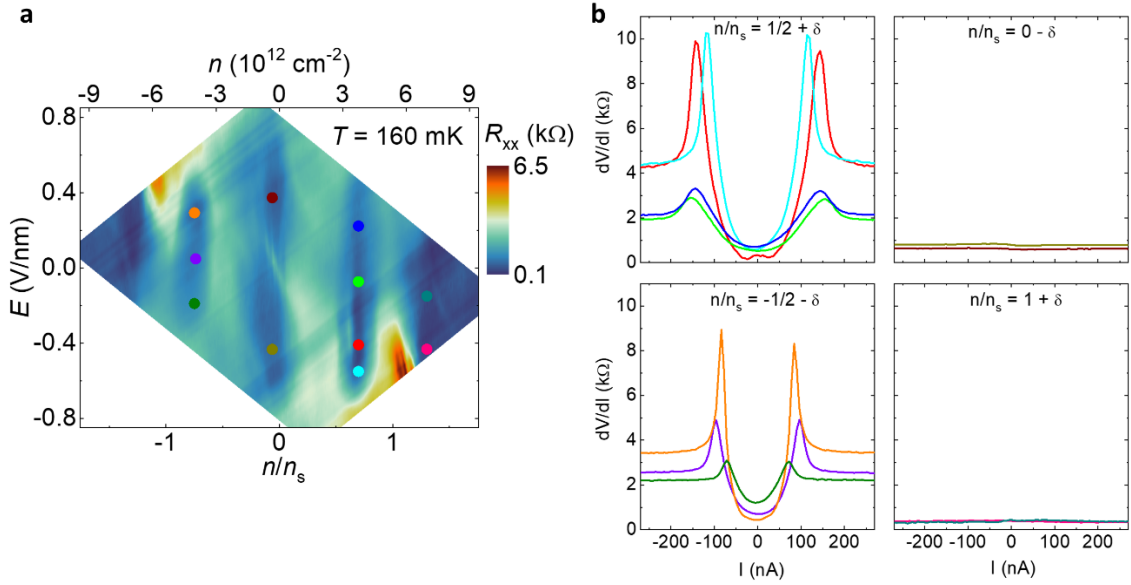

**Extended Data Fig. 3. d$V$/d$I$ vs $I$ at different $n/n_s$ and $E$. a**, $R_{xx}$ vs. $n/n_s$ and $E$ in the $\theta = 1.52°$ sample, the same data shown in Fig. 3a of the main text. The dots indicate points where d$V$/d$I$ vs. $I$ measurements were taken. **b**, d$V$/d$I$ vs. $I$ at $n/n_s = 1/2 + \delta$ (upper left), $0 - \delta$ (upper right), $-1/2 - \delta$ (lower left), and $1 + \delta$ (lower right). The line colors correspond to the dot colors in **a**. Two primary observations can be made from the data. First, regions near half filling show a critical current in the d$V$/d$I$ vs. $I$ at all $E$, while data acquired at other fillings show a d$V$/d$I$ that is insensitive to $I$. Second, the peaks in d$V$/d$I$ are generally most pronounced at larger $E$, in agreement with the theoretical picture that the critical temperature increases at finite $E$ for samples with a twist angle below the magic angle[24].

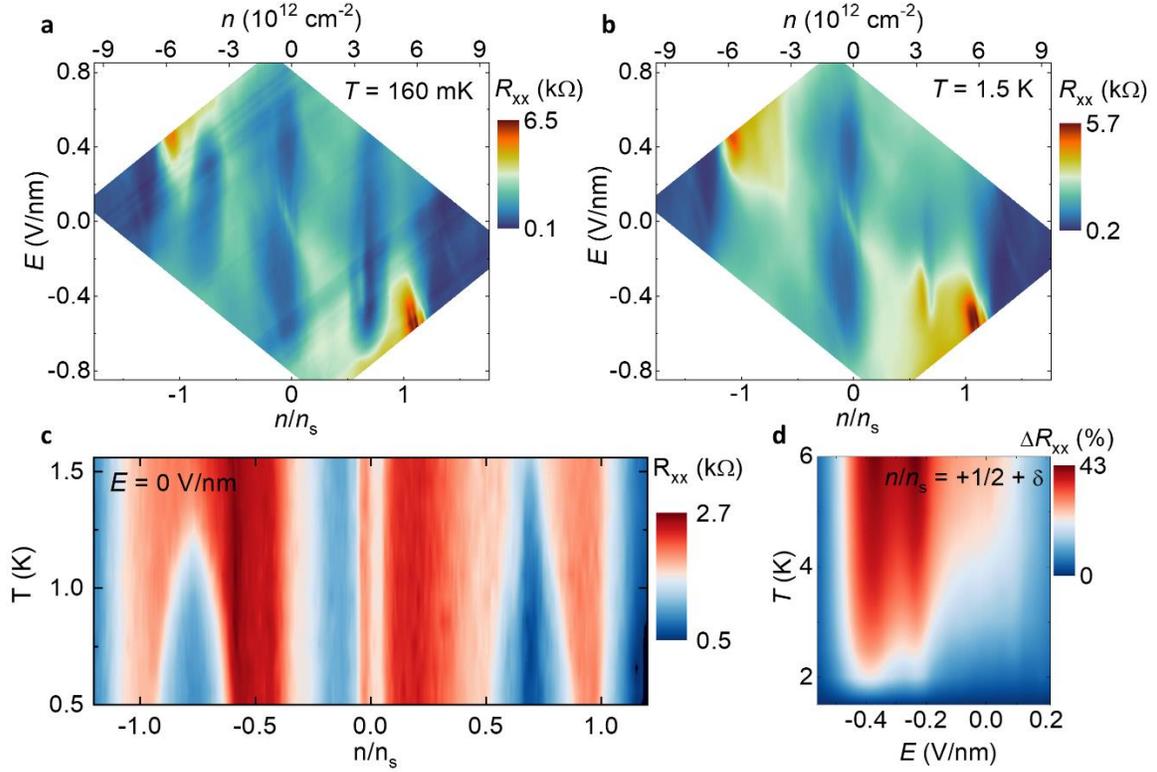

**Extended Data Fig. 4. Additional temperature dependence in $\theta = 1.52°$ ATQG. a, b**, $R_{xx}$ vs. $n/n_s$ and $E$ at $T = 160$ mK (**a**) and $T = 1.5$ K (**b**). At the lower temperature, $E$-dependent regions of low resistance appear at $n = \pm(n_s/2 + \delta)$ while all other regions remain insensitive to temperature between the two contour plots. **c**, $R_{xx}$ vs. $n/n_s$ and $T$ at $E = 0$ V/nm. Low resistance domes form on the higher density side of half filling on both the electron and hole side. Consistent with **a**, **b**, the data suggest a higher critical temperature on the electron side compared to the hole side. **d**, $\Delta R_{xx}$ vs. $E$ and $T$ at $n/n_s = 1/2 + \delta$ for $T$ between 1.5 K and 6 K, where $\Delta R_{xx}$ is the percentage change in $R_{xx}$ from the $T = 1.5$ K value. $R_{xx}$ changes more rapidly at larger $E$, suggesting an onset of correlations at higher temperatures with increasing $E$.

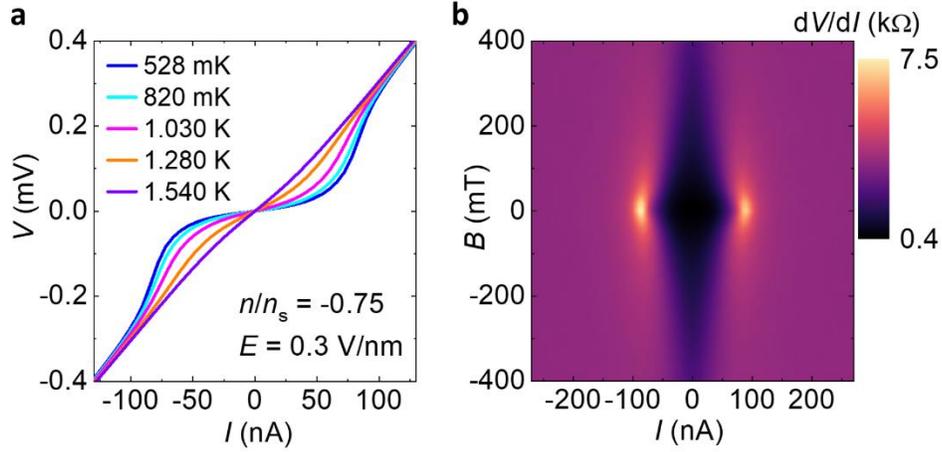

**Extended Data Fig. 5. Signatures of superconductivity on hole side. a**, $V$ vs. $I$ at $n/n_s = -0.75$ and $E = 0.3$ V/nm for temperatures between 528 mK and 1.540 K in the $\theta = 1.52°$ sample. At the highest temperature, the critical current behavior is extinguished and the characteristic becomes linear. **b**, $dV/dI$ vs $I$ and $B$ at the same $n/n_s$ and $E$ as in **a**. A small $B$-field destroys the nascent superconductivity, similar to the Fig. 4a data of the main text.

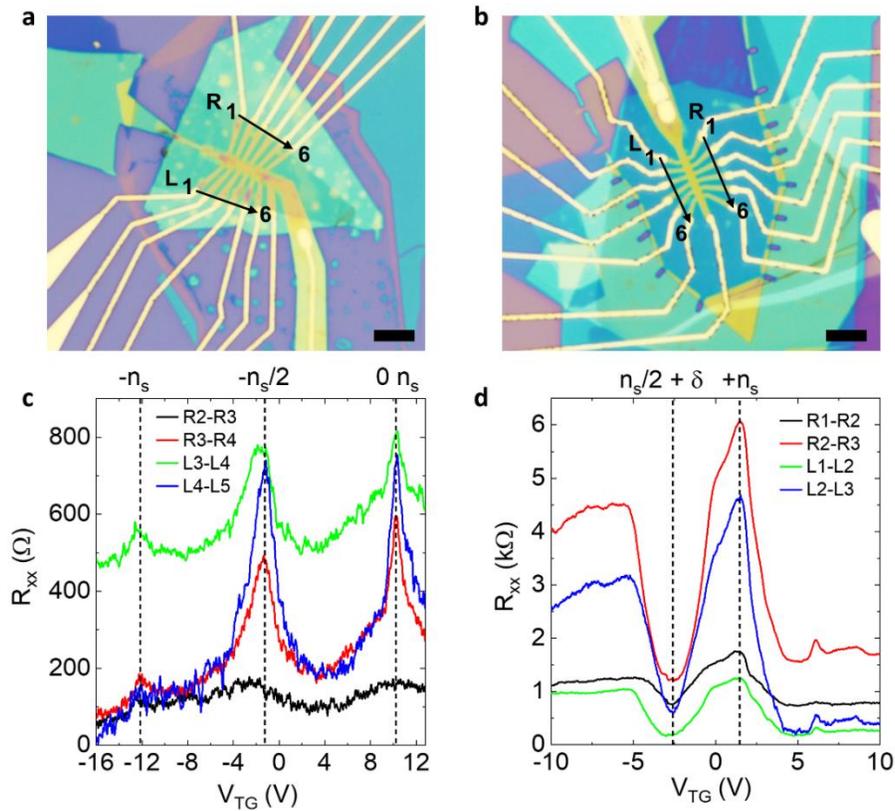

**Extended Data Fig. 6. Twist angle uniformity in the ATQG samples. a-b**, Optical micrographs of the $\theta = 1.96°$ (**a**) and $\theta = 1.52°$ (**b**) samples. The voltage probe contacts are labeled in each panel, and the scale bars are 5 μm. **c-d**, representative traces of $R_{xx}$ vs. $V_{TG}$ in the $\theta = 1.96°$ (**c**) and $\theta = 1.52°$ (**d**) samples using four different voltage probe contact pairs. For the $\theta = 1.96°$ sample, the $R_{xx}$ maxima corresponding to the band insulators at $0n_s$ and $-1n_s$ closely align, confirming a uniform twist angle to within ±0.01° throughout the channel. Similarly, in the $\theta = 1.52°$ sample, the positions of the $+1n_s$ $R_{xx}$ maxima and resistance dips at $n_s/2 + \delta$ that precede the superconducting state closely match across all contact pairs.

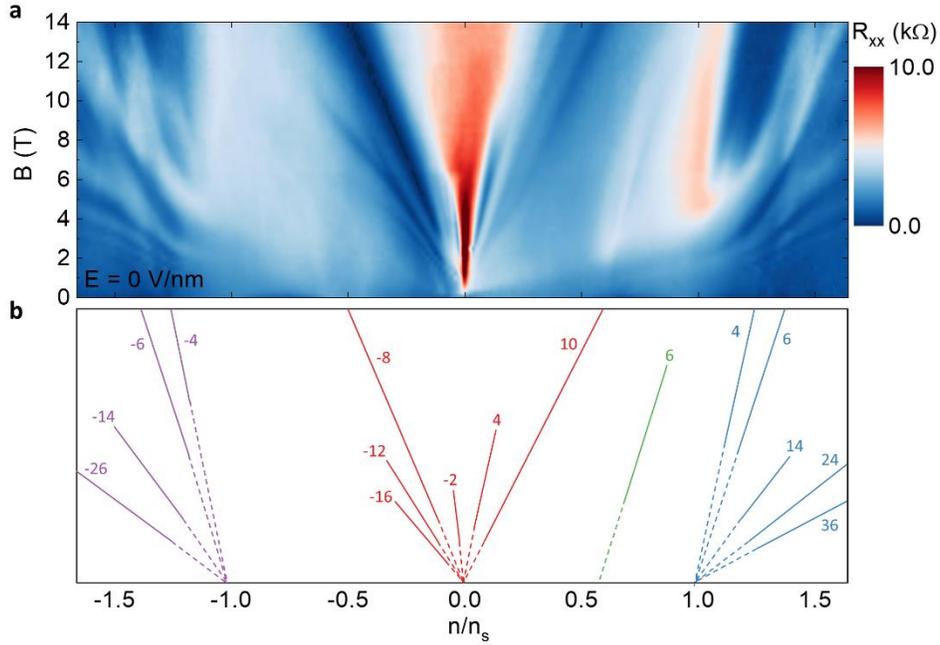

**Extended Data Fig. 7. Landau fans in $\theta = 1.52°$ ATQG. a**, $R_{xx}$ vs. $n/n_s$ and $B$ at $E = 0$ V/nm, showing quantum oscillations as described in the Methods section. **b**, Landau fans observed in panel **a**, along with the Landau level filling factor. The filling factor of 6 at $n = n_s/2$ is indicative of a band with Chern number 2, as discussed in the main text for the $\theta = 1.96°$ sample.

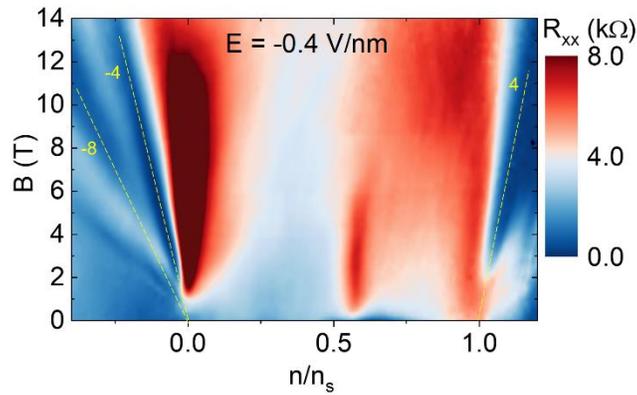

**Extended Data Fig. 8. Landau fans near nascent superconductivity.** $R_{xx}$ vs. $n/n_s$ and $B$ at $E = -0.4$ V/nm in the $\theta = 1.52°$ sample. The prominent Landau levels are labeled. Near half filling, the low resistance domes diminish at small $B$ values and a resistance maximum develops at the correlated insulator.